\documentclass[a4paper,twoside,twocolumn]{article}

\usepackage[top=1in, bottom=1in, left=0.8in, right=0.8in,columnsep=20pt]{geometry} % Document margins
\usepackage{booktabs} % Horizontal rules in tables
\usepackage{float} % Required for tables and figures in the multi-column environment - they need to be placed in specific locations with the [H] (e.g. \begin{table}[H])
\usepackage{hyperref} % For hyperlinks in the PDF

\usepackage{titlesec} % Allows customization of titles
\renewcommand\thesection{\Roman{section}} % Roman numerals for the sections
\renewcommand\thesubsection{\Roman{subsection}} % Roman numerals for subsections
\titleformat{\section}[block]{\large\scshape\centering}{\thesection.}{1em}{} % Change the look of the section titles
\titleformat{\subsection}[block]{\large}{\thesubsection.}{1em}{} % Change the look of the section titles

\usepackage{amsmath,amssymb,graphicx}

\title{\vspace{-15mm}\fontsize{18pt}{10pt}\textbf{Performance analysis of cone detection algorithms} \bigskip} % Article title

\author{
\medskip
\large
Letizia Mariotti \, and \, Nicholas Devaney\\[2mm] % Your name
\normalsize Applied Optics Group, School of Physics, National University of Ireland, Galway, Ireland \\ % Your institution
\normalsize \href{mailto:nicholas.devaney@nuigalway.ie}{nicholas.devaney@nuigalway.ie} % Your email address
\vspace{5mm}
}
\date{February 2015}

\begin{document}

\twocolumn[{
\maketitle 

\begin{abstract}
Many algorithms have been proposed to help clinicians evaluate cone density and spacing, as these may be related to the onset of retinal diseases. However, there has been no rigorous comparison of the performance of these algorithms. In addition, the performance of such algorithms is typically determined by comparison with human observers. Here we propose a technique to simulate realistic images of the cone mosaic. We use the simulated images to test the performance of two popular cone detection algorithms and we introduce an algorithm which is used by astronomers to detect stars in astronomical images. We use Free Response Operating Characteristic (FROC) curves to evaluate and compare the performance of the three algorithms. This allows us to optimize the performance of each algorithm. We observe that performance is significantly enhanced by up-sampling the images. We investigate the effect of noise and image quality on cone mosaic parameters estimated using the different algorithms, finding that the estimated regularity is the most sensitive parameter.
\end{abstract}

\bigskip

Link to the article: \url{http://www.opticsinfobase.org/abstract.cfm?msid=224577}

\bigskip
\bigskip
}]

\section{Introduction}

Thanks to the advent of Adaptive Optics (AO) in the field of retinal imaging (\cite{Liang1997}, for a review see~\cite{Hampson2008}), it has been possible for clinicians to detect and resolve individual cells in the retina in vivo, while previously this was possible only for tissues examined post mortem. The great advancement produced by the use of AO makes the early detection of retinal diseases and their follow up, as well as their early treatment, a real possibility~\cite{Carroll2013}. 

One important application of AO is the imaging of the photoreceptor mosaic, which is made up of cone and rod cells. It has been shown that the quality of a subject’s vision is strongly related to the characteristics of the photoreceptor mosaic~\cite{Yellott1983,Williams1983,Williams1987}. With images that have individual photoreceptors resolved, it is possible to determine features such as the cell spatial distribution and density. In order to identify the photoreceptor number and locations in an image, the first method used is the manual labelling, where an expert visually analyses the image and chooses the coordinates of every feature believed to be a cone or a rod~\cite{Putnam2005}. This method is still considered by many as the most reliable reference~\cite{Garrioch2012}. However, this approach is not desirable if the amount of data is large, as it is generally time and labour consuming, especially in patients suffering from retinal diseases. Moreover, the quality of the images is not always good enough for visual analysis, and the clinicians need to process them further in order to enhance their quality~\cite{Kitaguchi2007}. Automatic tools for the detection of cones and rods are necessary in order to spare time and work for the clinicians, and to have the possibility to increase the number of patients and images for more statistically significant results~\cite{Li2007}. 

Several algorithms for cone detection and counting have been described~\cite{Li2007,Xue2007,Chiu2013,Turpin2011,Mohammad2010,Loquin2011,Cooper2013}. As mentioned, the performance of detection methods has always been estimated by comparing with manual labelling operated by an expert or by the authors themselves. In some cases~\cite{Li2007,Xue2007}, the results of the detections are also compared with histological data~\cite{Curcio1990} to verify that there is a general agreement. Even if reliable, the manual counting results can vary because of the personal sensitivity of the observer, so observations by more than one physician may be needed to increase the reliability of the reference data~\cite{Lombardo2014}. We are not aware of any systematic study and comparison of the performances of the algorithms presented so far. In order to move towards a truly automated cone detection process, we believe that the algorithms should be tested in order to know the margins of their reliability when they are run in automated mode, i.e. without further manual corrections.

In this study, we propose to use synthetic cone images based on a real cone image in order to compare the performance of different cone detection codes. The simulation process has the advantage of having objective reference data to use as ``ground truth''. In this way, the reference data does not depend on the sensitivity of an observer and the amount of data that can be analysed is greatly increased, leading to more statistically significant results. In our performance analysis we consider not only the number of cones, but also the accuracy of the parameters retrieved from the detections other than the cone density, such as the regularity of the packing and the distance between the cones.

\section{Techniques}

In recent years many algorithms have been proposed to automate the detection of photoreceptor cells. The first cone counting algorithms, and still the most used so far, with some enhancements by users, are those developed by Li and Roorda~\cite{Li2007} and by Xue et al.~\cite{Xue2007}.

The Li and Roorda algorithm is based on the detection of local maxima in the image. The first step is to filter the image using a Gaussian low-pass filter, with the aim of removing high-frequency noise. The standard deviation of the Gaussian filter is set to 1 $\mu m$, which is half of the minimum possible cone separation (i.e. the mean nearest neighbour distance of the cones in the central fovea~\cite{Curcio1990}). After the filtering, the local maxima are found using the \textit{imregionalmax} built-in function of Matlab (Mathworks Inc). In the last step if two or more maxima are closer than the minimum cone separation their centroid is taken as the final location.

Xue et al. base their detection technique on an image histogram analysis. After applying a background subtraction, the image is divided in intensity ranges. Starting from the highest range, the algorithm searches the connected regions of pixels with intensity values which are in the range. The centroids of the connected regions are defined as the cone coordinates and if two or more coordinates occur closer than the minimum cone separation, their centroid is taken as the final location. A portion of the image with a size set by the user and surrounding each detected cone is excluded from later detection. This process is repeated for each intensity range, from the highest to the lowest. 

Chiu et al.~\cite{Chiu2013} recently proposed another algorithm. It starts by taking the local maxima as the initial cone detections. A portion of the image surrounding each cone is transformed from the Cartesian domain using the \textit{quasi-polar transform}, developed by the authors in a previous work~\cite{Chiu2012}. The resulting layered structures are segmented using graph theory and dynamic programming and then the segmentation boundaries are transformed back in the Cartesian domain. This process is iterated for all cones starting from the brightest values and excluding pixels belonging to other cones or segmentations. The entire process is repeated after deblurring the image using maximum likelihood blind deconvolution. The centroids of the segmented areas are the cone coordinates.

Some of the algorithms proposed later used correlation with a cone shape defined by the user~\cite{Turpin2011}, content-adaptive filtering in order to emphasize cell structure~\cite{Mohammad2010} and watershed by immersion for improved computational efficiency~\cite{Loquin2011}. Another way to measure the photoreceptor spatial density is to look at the image power spectrum and extract the features that correspond to the cell packing~\cite{Cooper2013}, but in this way it is not possible to retrieve the position of the individual cells. The presently proposed algorithms vary both in the degree of automation (i.e. the number of parameters that have to be set by the user) and in the performance of detection.

Among the cone detection algorithms, we chose to test only the algorithms of Li and Roorda, Xue et al. and Chiu et al., as they are the most commonly used and their approaches are radically different. In order to do that, we wrote the codes for the three algorithms according to the descriptions given in the original works. We also evaluated algorithms developed in other image processing fields that could be used for the purpose of cone counting. We initially tested two algorithms for feature extraction, the Harris and Stephen corner detection method~\cite{Harris1988} and the Scale Invariant Feature Transform~\cite{Lowe2004}. These two methods proved to be not suitable for our task, as the average cone shape is generally not sharp enough to be considered a corner and the algorithms detect local minima as well.

We tested instead an algorithm used in astronomy for the detection of stars in crowded fields. We used IRAF/DAOPHOT, a package for stellar photometry available within the IRAF data reduction and analysis environment for astronomy~\cite{Davis1994}, and in particular the utility task that searches for point sources, \textit{daofind}. The \textit{daofind} algorithm first convolves the input image with a Gaussian function having user-defined width, at each pixel this gives the amplitude of the least square best-fitting Gaussian. It then searches for local maxima in the convolved image whose amplitudes are greater than a detection threshold set by the user. The centres of these local maxima are the cone coordinates.

The final choice of these four algorithms is due to the following reasons; the Li and Roorda algorithm is simple and can in principle be used without changes for the analysis of images with different cone separation. On the other hand, it may miss detections in regions with low contrast (and typically low brightness) or densely packed cones. Xue et al. has the advantage that it detects cones at all intensity levels of the image, but if the deletion step is not done properly it may cause false detections. Chiu et al. is completely automated and does not require in principle the tuning of its parameters, but its main drawback is the computational complexity and the time required to run it. Daophot uses convolution, which is computationally expensive, but compared to the other algorithms it should not have a preference for the bright cones over the dim ones. Moreover, the coordinates are calculated with subpixel precision, and this could benefit the study of the cone arrangement.

\section{Method}

\begin{table*}[tb]
  \caption{\label{tab1} Parameters of the series of simulated images, with their mean and standard deviation in the last row. The mean density and percentage of hexagonal cells is compatible with values of previous works at corresponding retinal eccentricity~\cite{Curcio1990, Li2007, Park2013}.}
  \begin{center}
    \begin{tabular*}{\textwidth}{@{\extracolsep{\fill} } ccccc}
    \hline
    Image  & Cone density         &   Number of Cones        & Regularity (Percentage of     & Mean NND     \\
               &  (cones$/mm^2$) &  (boundaries excluded)  &  hexagonal cells)  &   ($\mu m$)  \\             
    \hline\noalign{\smallskip}
    1	& 25,933	& 3606	& 46.03	& 4.76 \\
	2	& 25,583	& 3557	& 47.17	& 4.82 \\
	3	& 25,691	& 3576	& 45.33	& 4.80 \\
    4	& 25,598	& 3566	& 45.54	& 4.81 \\
	5	& 25,805	& 3592	& 47.80	& 4.79 \\
	6	& 25,825	& 3596	& 45.13	& 4.78 \\
	7	& 25,607	& 3570	& 46.44	& 4.79 \\
	8	& 25,764	& 3588	& 48.10	& 4.80 \\
	9	& 25,663	& 3576	& 46.39	& 4.81 \\
	10	& 25,867	& 3600	& 45.14   & 4.77\\
    Mean $\pm$ St Dev & 25,734 $\pm$ 123	& 3583 $\pm$ 16	& 46.31 $\pm$ 1.09	& 4.79 $\pm$ 0.02\\
    \hline
    \end{tabular*}
  \end{center}
\end{table*}

\begin{figure}[!b]
\centerline{\includegraphics[width=.85\columnwidth]{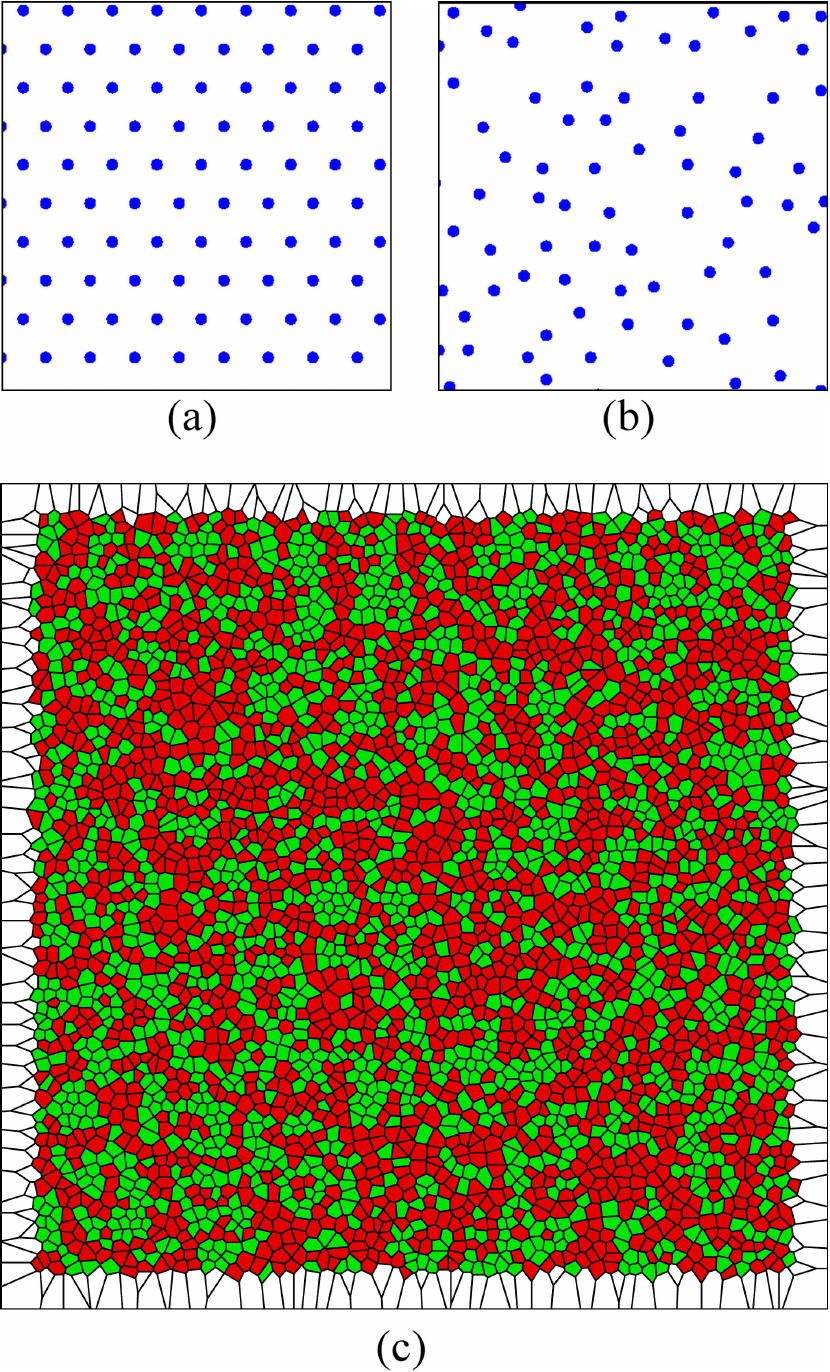}}
\caption{\label{fig1} (Color online) Detail of the grid of points with hexagonal arrangement (a) and of the same grid after the process of coordinate definition (b). The Voronoi diagram of the final arrangement of one of the simulated images (c). The diagram shows in light grey (green online) the cells with six sides, while all the other cells are coloured in dark grey (red online).}
\end{figure}

The retinal images used in this study were acquired using a commercial AO-assisted fundus imager, the rtx1 from Imagine Eyes~\cite{Lombardo2012b}. The pixel size on the retina is approximately 1.5 $\mu m$ for an emmetropic eye. We focused on our available images where the cones are best resolved by the imager, in particular images acquired at 2.5 degrees temporal from the foveal centre. 

We used as reference a 16-bit image of a healthy eye obtained from a series of 40 images that were processed for uneven illumination and registered with the procedure presented by Ramaswamy and Devaney~\cite{Ramaswamy2013}. We selected a portion of 256x256 pixels in the centre of the registered image without blood vessels, to be considered as our reference for the simulation of a cone image. 

We believe that in order to have a more realistic simulation, the cone shapes should account for non-integer centering. For this purpose, during the simulation process we used an image size three times bigger than the final size (768x768 pixels). In this way, the pixel sampling is increased from 1.5 $\mu m$ to 0.5 $\mu m$ and the cone centres in the final images are located on a grid that is three times finer than the pixel grid.

In order to simulate the cone packing, we started with an hexagonal grid of points, to which normally distributed random displacements are added. The Nyquist limit for a hexagonal array is $(\sqrt{3}d)^{-1}$ along one axis and $(2d)^{-1}$ along the other, where \textit{d} is the separation between the points. We note in passing that hexagonal sampling of the retina would be more efficient than rectangular sampling~\cite{Mersereau1979}. The cones are first simulated as uniform discs. Based on visual comparison with the reference image, the disc radius is set equal to quarter the mean cone separation. This relation will depend on the distance from the fovea of the simulated image. Any discs that overlap are merged, and their centroid taken as the new cone coordinate (Figure~\ref{fig1}a and~\ref{fig1}b). 

The regularity of the resulting packing is determined by Voronoi analysis and is defined as the percentage of hexagonal cells~\cite{Li2007}. The standard deviation of the random displacements can be chosen to reproduce the desired packing regularity: a bigger standard deviation will result in a lower percentage of hexagonal cells and so in a poorer regularity of the mosaic. A consequence of the merging of points is that the final density is different from the initial one, so its value has to be calculated again. Cells at image boundaries are excluded from this and all subsequent analyses (Figure~\ref{fig1}c).

We tuned our simulation so that the final density is compatible with the values provided by Curcio~\cite{Curcio1990} and the regularity of the mosaic has a value compatible with that found by Li and Roorda~\cite{Li2007} and Park~\cite{Park2013}. The synthetic cone images present a mean cone density of 25,734 cones$/ mm^2 $ and a mean percentage of hexagonal cells of $46.31 \%$, ranging from $45.13 \%$ to $48.10 \%$. The mean nearest neighbour distance, NND, has a mean value of 4.79 $\mu m$ (Table~\ref{tab1}).

Before using the described method, the first attempt at the simulation of the cone coordinates was made using a normal distribution truncated at a quarter the mean cone separation for the displacements, in order to avoid overlap of the cone discs. However, even though many values of the standard deviation of the distribution were tested, the regularity of the final mosaic was always much higher than the observed values. For this reason, we changed approach and used the merging of the cone discs instead. This agrees with results from previous studies, that suggest how the cone mosaic cannot be reproduced only by adding random deviations to a continuous hexagonal pattern~\cite{Lombardo2013c}. In general, cones seem to have a preferred arrangement that varies with the distance from the fovea, but the factors involved in the mosaic arrangement are not fully understood.

In order to assign the cone intensities we used as a starting point a smoothed version of the reference image, in order to reproduce the low-frequency brightness variation in the background and in the cones. The intensities were defined as values normally distributed around the values of the smoothed reference image at the cone positions. The standard deviation is chosen by comparing the histogram of cone brightness to that of the reference image (obtained by Li and Roorda algorithm).

\begin{figure}[!hb]
\centerline{\includegraphics[width=.85\columnwidth]{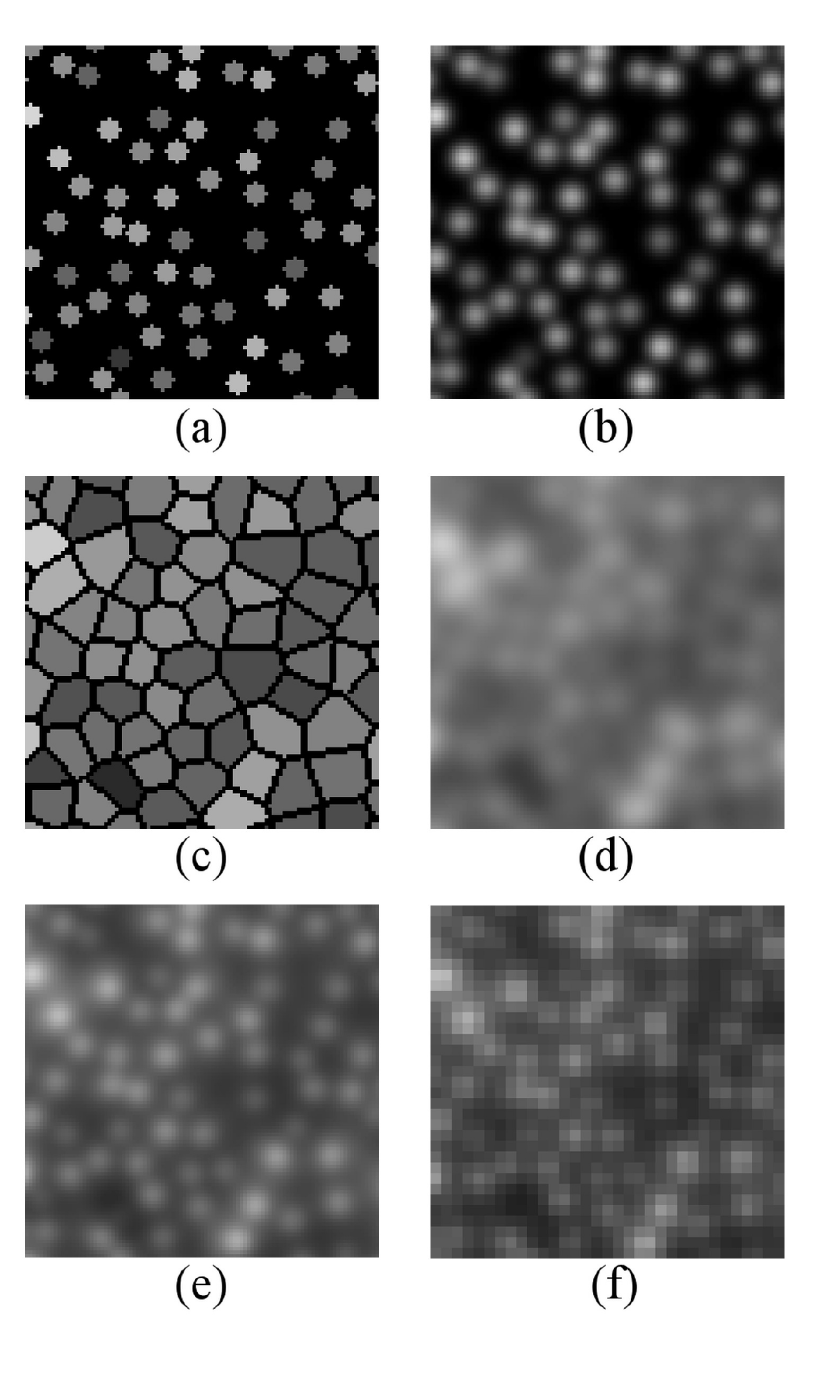}}
\caption{\label{fig2} In order to obtain the ``peaks'' image, circles are first placed (a) and then blurred with a Gaussian filter (b). The ``background'' image was obtained through the Gaussian blurring (d) of a Voronoi diagram (c). The two images (b) and (d) were then summed (e). The sum image was then rebinned by a factor three (f).}
\end{figure}

\begin{figure*}[tp]
\centerline{\includegraphics[width=\textwidth]{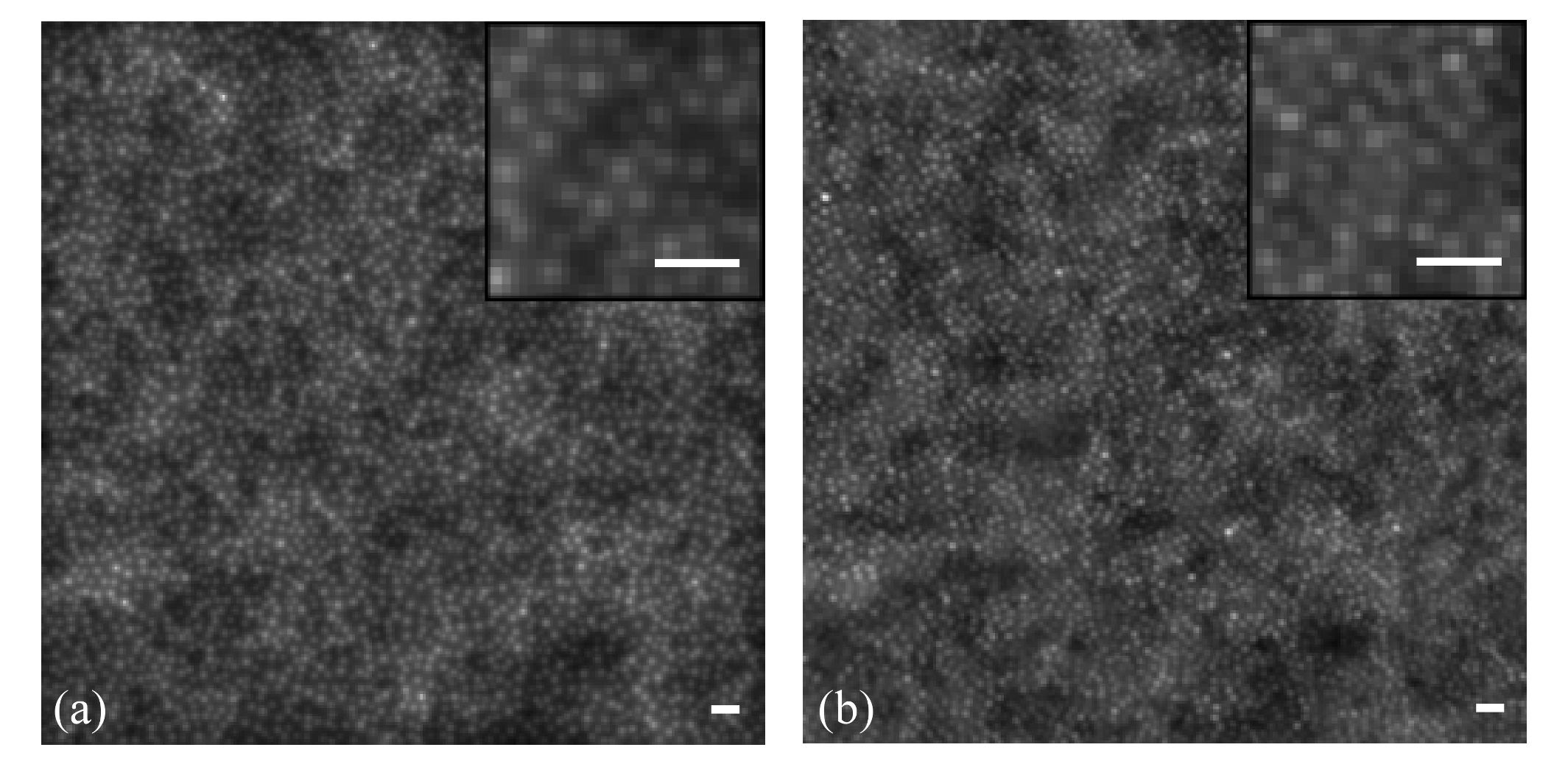}}
\caption{\label{fig3} An example of a synthetic image (a) obtained with the outlined process. The real image is shown in (b). The low-frequency variation of cone intensities follows the same pattern as the original image, but the coordinates and intensities of individual cones vary randomly. In the upper right corners details of the respective images are shown (scale bars are 15 $\mu m$). It can be noticed that the real image shows some patches with low brightness and no apparent presence of cones. It is not clear if this is due to illumination conditions or to the characteristics of the retina itself.}
\end{figure*}

Once the coordinates and the intensities of the cones are defined, the synthetic image was obtained by summing two images which we refer to as ``peaks'' and ``background'' images. The peaks image was obtained by placing circles at the cone positions with intensities obtained as described above and radius equal to a quarter of the mean cone separation, and then blurring them with a Gaussian filter, as done also in previous works~\cite{Xue2007}, with standard deviation equal to 0.75 times the radius of the circles. It is to be noted that, even if upsized by a factor 3, the sampling of the reference image is still too poor to reproduce a good disc shape. As background contribution, we used an image reproducing the Voronoi diagram, where every Voronoi cell has intensity equal to that of the cone at its centre. The Voronoi diagram image was also blurred with a Gaussian filter, broader than the one used for the peaks. The two images where subsequently summed. The resulting image was downsized by a factor of three using rebinning. Figure~\ref{fig2} illustrates the simulation process.  

The reason for using the sum of two different images as defined can be explained if we consider the images acquired by a fundus imager to be the result of convolution of the object, i.e. the photoreceptor layer, with a Point Spread Function (PSF) that presents both in-focus and out-of-focus contributions~\cite{Blanco2011}. If we consider that the light is reflected from the cone cells in two points at different depths, then the ``peaks'' and ``background'' images simulate the light reflected at two different depths and with different focusing, reproduced with the different size of the Gaussian filters. We found that the approach of smoothing the Voronoi diagram gives simulated images which are more similar to real images than other approaches such as smoothing the reference image or using broader Gaussians in the cone location. This was determined by evaluating the difference between the intensity histograms of images with different backgrounds with the Earth Mover's Distance metric~\cite{Rubner2000}. This metric was minimized in the case of the Voronoi background. 

Finally we added Gaussian noise to the image. The variance of the noise was estimated according to the method presented by Lee and Hoppel~\cite{Lee1989}, that considers the noise variance of an image as the minimum of the intensity variances calculated in blocks small enough to select only homogeneous areas of the image. We used for this the whole registered image, of which the reference image is a 256x256 pixels portion at the centre, as the whole image presents also blood vessels, which have homogeneous areas in their inner sections. Figure~\ref{fig3} shows the comparison between a simulated image and the real image used as reference.

Each counting algorithm (Li and Roorda, Xue et al., Chiu et al., DAOPHOT) was tested on these synthetic images. In each case, the Voronoi diagram of the detected cones was determined, in order to exclude the cells at the boundaries from the statistics and to compare the regularity of the mosaic. 

\section{Analysis}

In order to test the performance of the detection algorithms, we used a variant of the Receiver Operator Characteristic (ROC) curve. The ROC curve is widely used in medical imaging to estimate the performance of detection techniques~\cite{Barrett2004}. The ROC analysis was conceived for classification problems, i.e. when the detection process results in a positive/negative response, where positive means that a feature (e.g. a lesion) is present in the image and negative means that such feature is not present. In our case this is not applicable, as the task is to know how many cones are detected in the image and possibly if their position is correct. We therefore used a variant of the ROC analysis, the Free-response Receiver Operating Characteristic (FROC) curve.

The FROC analysis was introduced to integrate the localization task into the ROC analysis of data in the presence of multiple lesions~\cite{Bunch1977}. When the detections are compared with ground truth data, they are divided into two types: true positive (TP), when a detection is within a specified tolerance from a true lesion, and false positive (FP) otherwise. The results are summarized by a plot of the fraction of detected lesions (TP over number of real lesions) against the mean number of FP per image, obtained for different levels of confidence. The horizontal axis of the FROC curves is not normalized, as it can extend to an arbitrary number of FP. Each curve gives the detection performance for a detection algorithm (or ``observer''). The perfect observer curve is a straight line that lies on the vertical axis from 0 to 1, where all the true lesions are detected and there are no FP. In comparing the performance of different observers, we can say that one observer is better than another if its curve is closer to the vertical axis and/or is higher. 

In our study, the observers are represented by the detection algorithms. For each algorithm, a threshold (i.e. confidence level) was defined and varying the threshold traces a curve in the FROC plane. In the Li and Roorda algorithm the threshold parameter is the value of the dimmest local maxima that are found after applying the \textit{imregionalmax} function. In Xue et al., the threshold is the peak value of the dimmest cone, and in the original version of the algorithm this was set by the user. In DAOPHOT, the threshold is the amplitude of the dimmest local maxima in the convolved image. Chiu et al., compared with the other algorithms, does not have a parameter that can be straightforwardly chosen as a threshold. But as the algorithm, similarly to Xue et al., has a preference for bright cones over dim cones, we set the same threshold that we used for Xue et al., i.e. the peak value of the dimmest cone. The codes of the four algorithms were run by varying the threshold from the strictest one, with less detections, to the most relaxed one, with more detections.

Each of the four methods uses its own parameters besides the threshold value. The only parameter in Li and Roorda is the standard deviation of the Gaussian low-pass filter, but as it corresponds to the minimum cone separation this has a fixed value. The Xue et al. parameters are the standard deviation of the Gaussian filter that is applied for the estimation of the background, the number of intensity ranges and the size of the mask used to delete the detected cones. The Chiu et al. parameters were empirically found by the authors in the roginal work and were not modified here. In DAOPHOT the only parameter is the standard deviation of the Gaussian used for the convolution. Before comparing them, we run Xue et al. and DAOPHOT algorithms with different values of the parameters, to find the combination of values that results in the best performance. This was done by tracing an FROC curve for each parameter combination of each code and choosing the parameter values that gave the highest curve. Contrary to the ROC analysis, at present there is not a universally accepted method for fitting FROC curves nor a single index that summarises the overall performance~\cite{Bandos2009}. Two of the criteria commonly used to evaluate the detection algorithms are the calculation of the area under the curve up to a certain FP value~\cite{Strickland2002} or the TP fraction value at a fixed FP number~\cite{Chakraborty1986}. We use the latter in this work. 

\begin{figure}[bt]
\centerline{\includegraphics[width=.85\columnwidth]{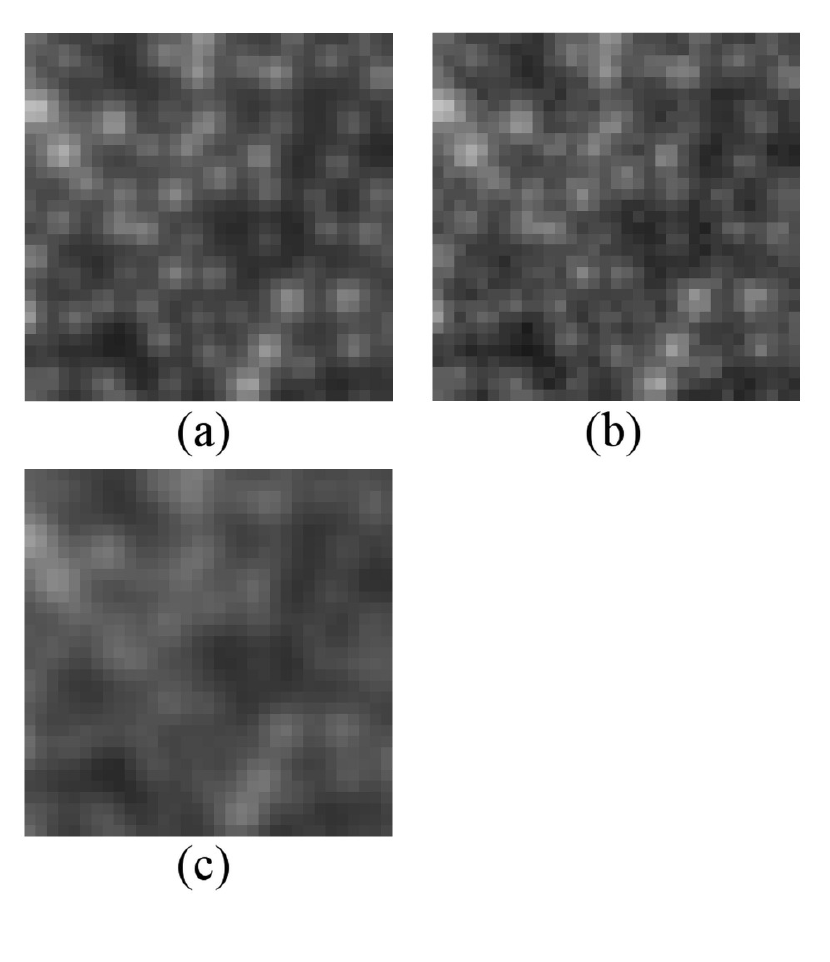}}
\caption{\label{fig4} Detail of a simulated image (a), detail of the same image with 5 times the original noise (b) and with Gaussian blur filter of 1 pixel standard deviation (c). }
\end{figure}

A problem that is not usually assessed in cone detection studies is how the quality of the images affects the detection. Registration and summing of frames is performed in order to obtain a final image with increased signal-to-noise ratio and quality~\cite{Ramaswamy2013}. However, the final image can present poor quality, especially in the case of diseased eyes~\cite{Lombardo2013b}. We therefore simulated two series of images with decreasing quality by applying a Gaussian blur of increasing width (with standard deviations of 0, 0.25, 0.5, 0.75 and 1 pixel) and by adding stronger noise (with standard deviation from 1 to 5 times the value retrieved from the reference image) to one image with the same features as the previously simulated images (Figure~\ref{fig4}). 

While DAOPHOT uses interpolation to retrieve the cone positions with sub-pixel accuracy, the cones found by the Li and Roorda and the Xue et al. algorithms have integer coordinates for the majority of the detections. Li and Roorda and Xue et al. detections, in fact, have non-integer coordinates only when the centroids are calculated for connected regions that are non-symmetrical with respect to the pixel grid, and this is not usually the case. Chiu et al. uses centroids of connected regions as well, but it is more common that the connected areas are non-symmetrical compared to the other two algorithms.

For this reason, we also performed the whole detection process after increasing the size of the images by a factor 2 and by a factor 3 using cubic interpolation~\cite{Lombardo2013c}, in order to increase the sampling to 0.75 $\mu m$ and 0.5 $\mu m$ per pixel and see how this affects the cone localization. Even if the resizing process does not add information to the images, we performed it because we wanted to determine if the accuracy of the localization process has a significant influence on the calculation of those image parameters that are position dependent, such as the percentage of hexagonal cells and the mean NND.

The benefit of using the FROC analysis is not only to have a method to compare the algorithms, but also to have an objective way to determine the best operating point along the curve, i.e. threshold value, at which the detection operates. As the horizontal axis is unbound, in clinical practice FROC curves are usually displayed only over the FP range that is considered of clinical interest, with the choice of a maximum acceptable value of FP. The cone density varies significantly even between individual healthy subjects  and at the eccentricity studied by us, 2.5 degrees temporal, the coefficient of variation can reach $15 \%$~\cite{Curcio1990,Lombardo2013b}. Therefore, we considered 500 FP (approximately to $15 \%$ of the mean number of cones per image) as the maximum acceptable FP value, determining that if the FP exceeds this value then the error is too large to be useful. The operating point was chosen similarly as for the ROC analysis by selecting the point on the curve closest to the upper left corner (0,1).

We therefore run four algorithms with the threshold at the operating point for every image, and then calculate the cone parameters (cone density, percentage of hexagonal voronois and mean NND) as if the results were not corrected by human intervention, i.e. considering all the detections as TP. This was done to test the reliability of the results of the algorithms if they are run in automated mode.

\section{Results}

\begin{figure}[t!b]
\centerline{\includegraphics[width=.96\columnwidth]{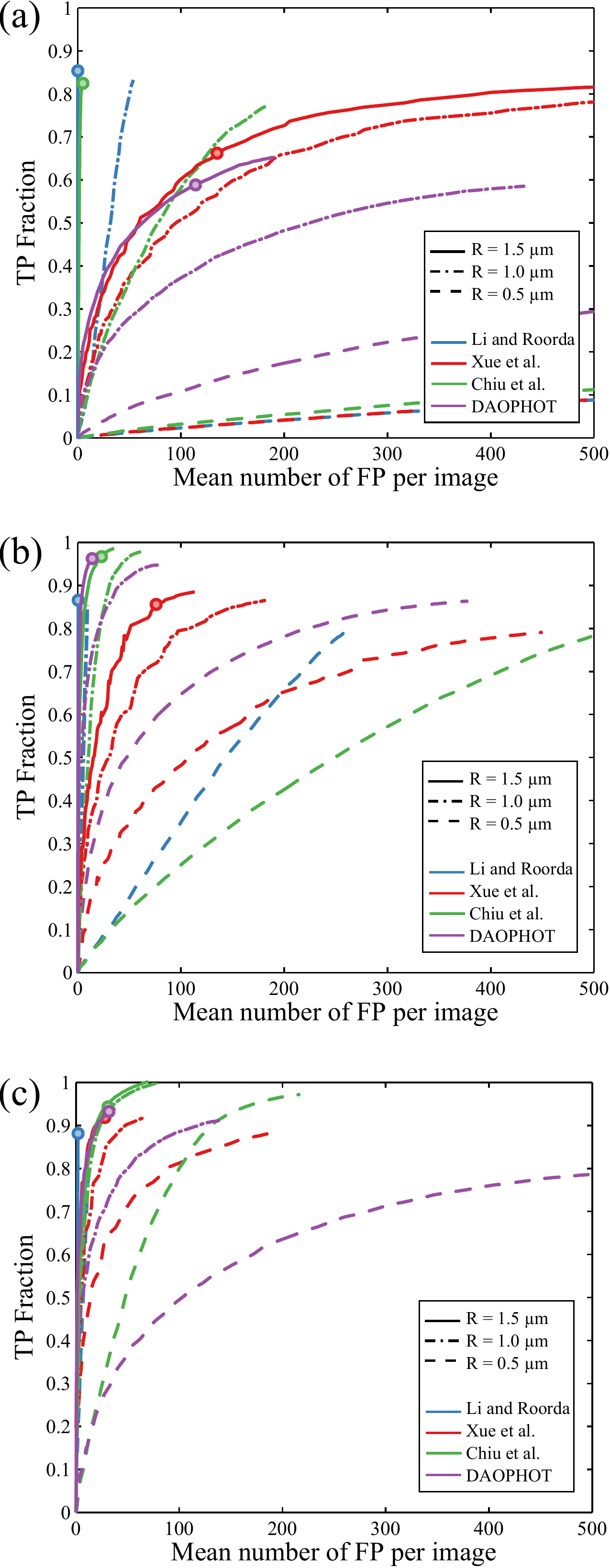}}
\caption{\label{fig5}(Color online) FROC curves for ten images without resizing (a), resized by two (b) and resized by three (c). Each plot displays the curves for the three tolerance radii (solid, dash-dot and dashed line) and for the four algorithms (blue, red, green and purple colours). The operating points are marked with circles on the solid lines.}
\end{figure}

\begin{table*}[tb]
  \caption{\label{tab2} Mean values and standard deviations of the parameters percentage differences for the detections on the images compared to the values presented in Table~\ref{tab1}.}
  \begin{center}
    \begin{tabular*}{\textwidth}{@{\extracolsep{\fill} } lccc}
    \hline
	 	& Cone density & Regularity & Mean NND\\
	\hline
	\textbf{Resize x1} & & & \\
	Li and Roorda   & $-13.83 \pm 0.92$ & $-10.68 \pm 2.81$ &	$5.69 \pm 0.53$ \\
	Xue et al.	        & $-28.00 \pm 3.68$ &	 $-20.59 \pm 2.50$	&  $9.32 \pm 2.65$ \\
	Chiu et al.	        & $-17.09 \pm 0.54$ &	 $-7.24   \pm 2.48$	&  $12.75 \pm 0.38$ \\
	DAOPHOT	        & $-37.41 \pm 2.00$ & $-18.35 \pm 3.10$	&  $10.80 \pm 1.48$\\
	\hline \noalign{\smallskip}	
	\textbf{Resize x2} & & & \\
	Li and Roorda	& $-12.88 \pm 0.86$ & $-8.83 \pm 2.19$ & 	$5.60 \pm 0.60$ \\
	Xue et al.	        & $-11.57 \pm 1.30$ & $-8.71 \pm 2.54$ &    $7.19 \pm 0.78$  \\
	Chiu et al.	        & $-2.68   \pm 0.64$ &	 $-0.58  \pm 1.62$	&  $4.67 \pm 0.34$ \\
	DAOPHOT	        & $-2.84   \pm 0.23$ &	 $-4.08 \pm 2.19$ &	$-1.85 \pm 0.35$ \\
    \hline \noalign{\smallskip}
    \textbf{Resize x3} & & & \\
	Li and Roorda   & $-11.39 \pm 0.66$ &$-8.48 \pm 2.39$	 & $4.78 \pm 0.49$ \\
	Xue et al.	        & $-7.32 \pm 0.73$ 	&$-4.91 \pm 1.64$ 	&  $6.36 \pm 0.39$ \\
	Chiu et al.	        & $-3.19 \pm 1.78$ &$-2.98  \pm 1.85$	&  $2.03 \pm 0.55$ \\
	DAOPHOT	        & $-5.18 \pm 0.35$	&$-6.93 \pm 2.42$	&  $-2.56 \pm 0.40$\\
    \hline
    \end{tabular*}
  \end{center}
\end{table*}

\begin{figure*}[t!b]
\centerline{\includegraphics[width=\textwidth]{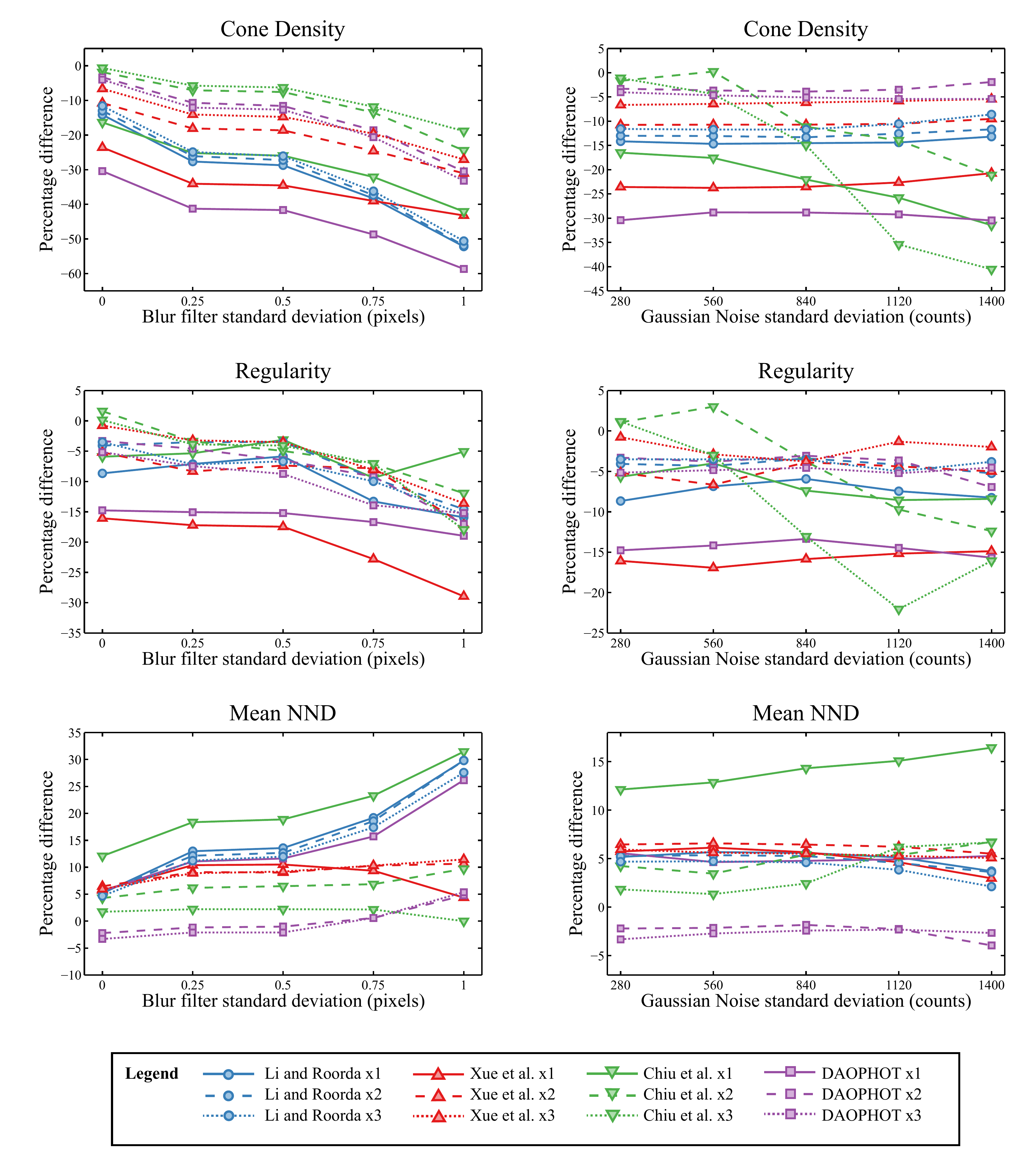}}
\caption{\label{fig6} (Color online) Cone parameters for increasing blur and noise. The resizing used is shown with different lines, the algorithms are marked with different colours and markers.}
\end{figure*}

\begin{figure}[t!b]
\centerline{\includegraphics[width=.96\columnwidth]{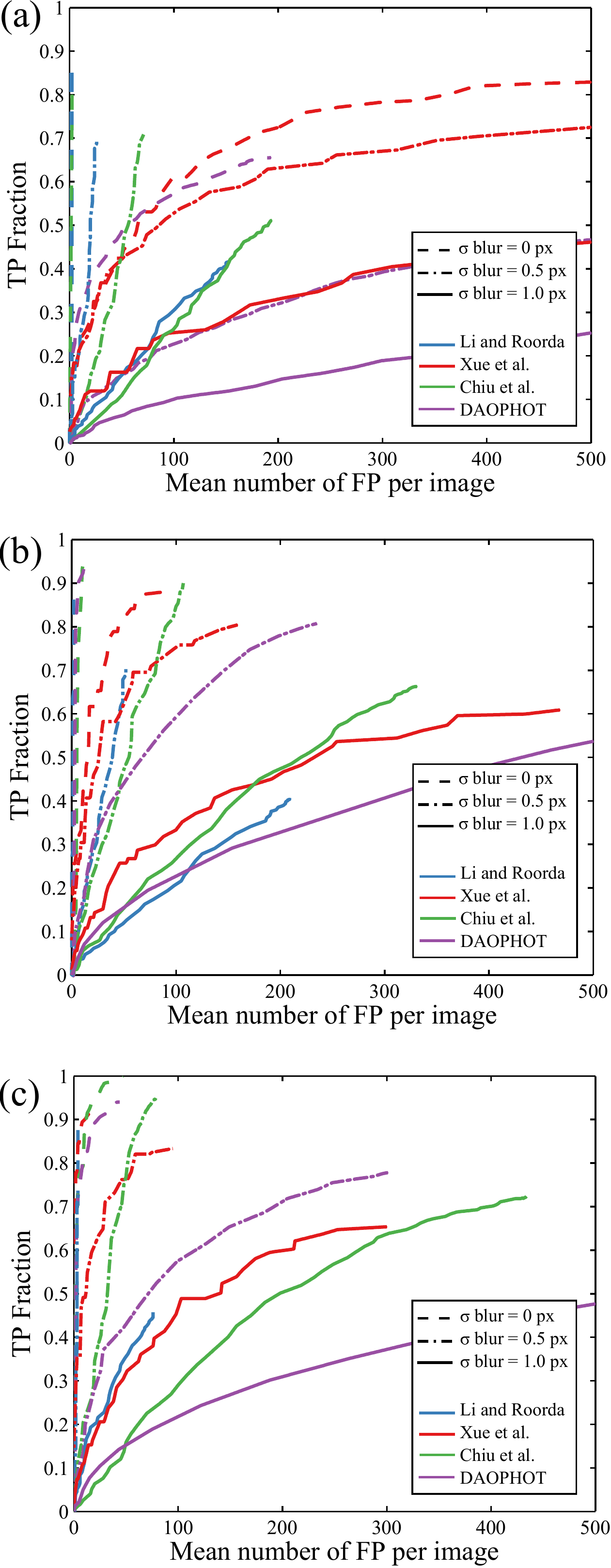}}
\caption{\label{fig7}(Color online) FROC curves for one image with increasing blur without resizing (a), resized by two (b) and resized by three (c). The tolerance radius used is 1.5 $\mu m$. Each plot displays the curves for three standard deviations of the Gaussian filter applied on the same image (dashed, dash-dot and solid line) and for the four algorithms (blue, red, green and purple colours).}
\end{figure}

In order to trace the FROC curves, we used three values for the radius of the tolerance region for the discrimination between TP and FP; 0.5, 1.0 and 1.5 $\mu m$, that correspond to 0.33, 0.67 and 1.0 pixel on the original size images. Figures~\ref{fig5}a,~\ref{fig5}b and~\ref{fig5}c show the FROC curves of the four algorithms averaged over 10 simulated images. 

As expected, the overall performance for all of the algorithms increases as the tolerance radius is relaxed. Li and Roorda is the algorithm that, according to the definition of FROC curve, performs best, with a curve very close to the vertical axis if exact localization is not required. Nonetheless, even if the algorithm gives almost no FP, the percentage of correctly detected cones in our case reaches $85 \%$ if the images are not resized. With resizing, the percentage slightly increases but does not exceed $90 \%$ in any case. With images at their original size, the operating point of the Xue et al. algorithm stops at $66 \%$ of TP, with a number of FP that corresponds to $4 \%$ of the real cones. However, the performance increases steadily with the resizing factor, as the TP fraction goes from $85 \%$ with a resizing of two to $92 \%$ with a resizing of three, with FP percentages that are respectively of $2 \%$ and $0.7 \%$. The Chiu et al. performance on original size images is similar to Li and Roorda, with $82 \%$ TP and almost no FP. After resizing is applied, the TP fraction increases to $97 \%$ and $94 \%$ respectively, with a FP percentage less than $1 \%$ in both cases. DAOPHOT performance in the first case is similar to Xue et al., with a TP fraction of $59 \%$ and a FP percentage of $3 \%$. For resized images, DAOPHOT performance at the operating point is practically identical to Chiu et al.

The cone parameters as calculated from the detections were compared to the real values, i.e. the values used in the simulation and presented in Table~\ref{tab1}. Table~\ref{tab2} shows the means and standard deviations of the percentage differences of the parameters. 

The cone density is underestimated by all the algorithms. The best overall estimates are achieved by Chiu et al. and DAOPHOT after doubling the size of the images, with a mean of $-2.68 \%$ and $-2.84 \%$ difference from the real value. If the images are resized, Xue et al., Chiu et al. and DAOPHOT achieve better estimates of the cone density than Li and Roorda, for which the mean percentage difference is greater than $10 \%$ in both cases. The regularity of the cone mosaic is the parameter that has the overall biggest uncertainty in its calculation, with standard deviations from $1.62 \%$ to $3.10 \%$. As for the cone density, the regularity is underestimated in every case, with a difference that goes from $-0.58 \%$ for Chiu et al. with resizing 2 to $-20 \%$ for Xue et al. with no resizing. The mean NND, on the other hand, is generally overestimated, with the exception of DAOPHOT. Compared to the corresponding estimates of cone density and regularity for the same algorithm, NND estimates are usually more accurate, with a difference that does not exceed $13 \%$ in any case.

Figure~\ref{fig6} shows how the quality of the images (noise and blur) affects the calculation of the cone parameters. The plots display the percentage difference of the parameters against the standard deviation of the Gaussian noise (in counts per pixel) and the standard deviation of the Gaussian filter used for the blurring (in pixels). The amount of noise in the image, even if increased to 5 times the value retrieved from the original images, does not significantly affect the results, with the exception of Chiu et al. The amount of blur, on the other hand, has a greater effect on the detections for all the algorithms. In the worst case, in fact, only $40 \%$ of the cones are detected and the mean NND peaks at $30 \%$ difference from the real value.

Figures~\ref{fig7}a,~\ref{fig7}b and~\ref{fig7}c show the FROC curves of the four algorithms for the same image with gradually decreasing quality (as we saw that the noise effect is limited, we include only the FROC curves for increasing levels of blur). The radius of the tolerance region corresponds to the most relaxed of the radii used in Figure~\ref{fig5} (1.5 $\mu m$). All of the algorithms show a significant decrease in their performance as the quality of the image worsens, leading to less detection of TP as well as more FP.

\section{Discussion}

In this study, we developed a method for the simulation of AO corrected images of cone photoreceptors and we used the simulated images to test the performance of four automated cone detection algorithms. 

If the simulated images are analyzed at their original size, then the Li and Roorda algorithm has the best results and it is, together with Chiu et al., the only one with an acceptable performance, with $85\%$ of the cones detected. It can be seen that the process of increasing the size of the images by factors of two and three improves the performance of Xue et al., Chiu et al. and DAOPHOT algorithms, even if no new information is added. We believe that the main cause for this change is the better sampling of the images. In Xue et al. the fact that the cones are better sampled allows a more precise deletion in the deletion step, reducing the detection of FP at the borders of the deleted areas. The best performance reached by the algorithm, $96\%$ of TP, is close to the value given in the original study, $97.7\%$~\cite{Xue2007}. Chiu et al. also benefits from the improved sampling of the images, as the segmentation of the cones is more precise. In the case of DAOPHOT, a Gaussian function is used to fit the cones. The Gaussian function is the best choice for well-sampled images, i.e. PSF with a full width half maximum of at least 2.5 pixels~\cite{Davis1994}, but this is approximately the size of the cones with the original sampling. Doubling the size of the images is already sufficient to reach a very good performance with DAOPHOT, and it is not improved by further increasing the size.

Even if the FROC curves are a useful way to visualize the overall performance of the algorithms, we are more interested in the accuracy of the cone packing parameters (Table~\ref{tab2}), as these represent the information actually used by clinicians. Without the discrimination between TP and FP, the most noticeable consequence is that the number of FP detections can compensate for an equal number of missed TP, leading to a better estimate of the cone density with an actually poorer detection performance. For this reason, the choice of an operating point is very important as it actually sets a limit to the number of FP and to apparently good but actually incorrect density estimates. From Table~\ref{tab2}, we can see that density is always underestimated. The fact that the Li and Roorda algorithm incurs fewer FP but also has fewer TP than the other algorithms at their best leads to worse estimates of cone density and mosaic regularity.

The mean NND results improve with the resizing, as the increased sampling allows a better determination of cone positions. It is also to be noticed that a better estimate of one parameter, i.e. mean NND, does not necessarily correspond to an improvement in other parameters such as regularity. 

As in the case of cone density, the percentage of hexagonal Voronoi cells is underestimated in every case, as the undetected cones cause a greater deviation from the hexagonal arrangement. Compared to density and NND, regularity shows no significant improvement with resizing from double to triple. Together with this, it is also the parameter with the largest uncertainty. These reflections suggest that the Voronoi analysis results are less stable than other metrics, and so their reliability needs to be thoroughly considered when used clinically, as noticed also in recent works~\cite{Lombardo2014}.  

It is noticed that while doubling the size of the images provides a major improvement in the parameter estimates, resizing by a factor of three leads to further improvement in only a limited number of cases. Moreover, the best performance for all three parameters is achieved using the resize factor of two, the DAOPHOT algorithm for cone density and mean NND and Chiu et al. for the regularity. This suggests that in order to achieve a significant improvement in the parameter accuracy an excessive increase of the sampling, which would require more computational time as well as computer memory, is not needed. Chiu et al. is the only algorithm that detects all the cones if the threshold is the most relaxed. It is to be noted though that this is possible at the expense of the speed of the detection process, as this algorithm is the most complex of those analyzed. 

In Figure~\ref{fig6}, it can be seen that the addition of noise affects significantly only the results of Chiu et al. algorithm. Li and Roorda use a low-pass filter to remove the high-frequency noise and this step proves to be effective, as its results remain substantially constant as the noise increases. For this reason, we believe that the dependence of the performance of Chiu et al. on the noise can be stabilized in a similar way by using low-pass filtering as the first step in the detection process. Even if the two other algorithms show a small decline in their performance too, a preliminary low-pass filtering can in principle be added to both of them to further limit the effect of noise. 

By contrast, the deterioration due to the blurring notably worsens the performance of the algorithms. All of the algorithms are affected in a similar way by blurring. Xue et al., Chiu et al. and DAOPHOT seem to produce inconsistent results, with an improvement in the NND measurement while density and regularity worsen. This could be caused by a significant number of FP. 

In the case of the Li and Roorda algorithm, the dependence of the performance on the level of blur in the image could explain why in our study the percentage of detections never exceeds $90\%$, while in the original work the agreement between the manual and the automated labelling is between $93\%$ and $96\%$~\cite{Li2007}. It is likely that this difference is due to the quality of the images used. Our images were simulated using as reference images acquired with a commercial AO fundus camera. Li and Roorda, on the other hand, used images acquired in an university laboratory both with flood-illuminated and scanning laser ophthalmoscope AO systems, that generally achieve better contrast~\cite{Hampson2008}. Therefore, we can say that the quality of an image data set, which can depend both on the condition of the subject’s eye and on the imaging system used, has a significant impact on the performance of the cone detection algorithms. 

Finally, it is to be noted that Li and Roorda and Chiu et al. are the only algorithms that do not have a parameter that accounts for the size of the cones. Because of this, they can in principle be used without modifications for images of cones acquired at different distances from the fovea. On the other hand, Xue et al. and DAOPHOT have size dependent parameters, respectively the size of the mask that excludes the detected cones and the radius of the Gaussian fitting function, that would need to be optimized every time. 

\section{Conclusions}

We have developed a method for the simulation of retinal images of cones in the parafoveal region and we used it to reproduce images acquired with a commercial AO-assisted fundus camera. The simulation data were used to test and compare the performance of three automated cone detection algorithms. We introduced the use of FROC analysis to optimize the algorithm parameters and to determine the operating point for each algorithm. The performance of the algorithms was then compared considering both the TP and FP detections and the estimates of the cone parameters (density, packing regularity and mean NND). 

We found that the spatial sampling of the images, even using resizing of the recorded images, has a significant impact on the performance of the algorithms, but also that an excessive upsizing does not improve substantially the measurements. It is suggested that the image sampling should always be provided when presenting results obtained using these algorithms.

We saw that the percentage of hexagonal Voronoi cells is the parameter which is most affected by errors in cone detection, and for this reason the combined measurements of more parameters could be a better choice in order to characterize different retinal regions and the retinas of different subjects~\cite{Lombardo2014b}.

We studied moreover how the performance was affected by the introduction of a variable quantity of noise or blur, finding that the level of blur significantly affects the detections and the derived parameters.    
 
As the algorithms were tested when used automatically, we believe that the results of this paper can be taken as reference for their reliability and accuracy for clinicians who want to perform cone detection without further manual supervision. The detection of cones in diseased eyes or in low-quality images, on the other hand, still requires accurate supervision, especially when the difference with healthy eyes is subtle~\cite{Lombardo2014b}, and needs to be addressed in future studies. 

\section{Acknowledgments}

The authors would like to thank the Irish Research Council (IRC) (GOIPG/2013/775) for financial support. We thank also Dr. Marco Lombardo (Rome, Italy) for providing the image data set and for his valuable comments and revision of the manuscript.


\begin{thebibliography}{10}
\newcommand{\enquote}[1]{``#1''}

\bibitem{Liang1997}
J.~Liang, D.~R. Williams, and D.~T. Miller, \enquote{Supernormal vision and
  high-resolution retinal imaging through adaptive optics,} J. Opt. Soc. Am. A \textbf{14},
  2884--2892 (1997).

\bibitem{Hampson2008}
K.~M. Hampson, \enquote{Adaptive optics and vision,} Journal of Modern Optics
  \textbf{55}, 3423--3465 (2008).

\bibitem{Carroll2013}
J.~Carroll, D.~Kay, D.~Scoles, A.~Dubra, and M.~Lombardo, \enquote{Adaptive
  optics retinal imaging - clinical opportunities and challenges,} Current Eye
  Research \textbf{38}, 709--721 (2013).

\bibitem{Williams1983}
D.~Williams and R.~Collier, \enquote{Consequences of spatial sampling by a
  human photoreceptor mosaic,} Science \textbf{221}, 385--387 (1983).

\bibitem{Williams1987}
D.~R. Williams and N.~J. Coletta, \enquote{Cone spacing and the visual
  resolution limit,} J. Opt. Soc. Am. A \textbf{4}, 1514--1523 (1987).

\bibitem{Yellott1983}
J.~I. Yellott, \enquote{Spectral consequences of photoreceptor sampling in the
  rhesus retina,} Science \textbf{221}, 382--385 (1983).

\bibitem{Putnam2005}
N.~M. Putnam, H.~J. Hofer, N.~Doble, L.~Chen, J.~Carroll, and D.~R. Williams,
  \enquote{The locus of fixation and the foveal cone mosaic,} Journal of Vision
  \textbf{5(7)}, 3 (2005). 

\bibitem{Garrioch2012}
R.~Garrioch, C.~Langlo, A.~M. Dubis, R.~F. Cooper, A.~Dubra, and J.~Carroll,
  \enquote{Repeatability of in vivo parafoveal cone density and spacing
  measurements,} Optometry and Vision Science \textbf{89}, 632--643 (2012).

\bibitem{Kitaguchi2007}
Y.~Kitaguchi, K.~Bessho, T.~Yamaguchi, N.~Nakazawa, T.~Mihashi, and
  T.~Fujikado, \enquote{In vivo measurements of cone photoreceptor spacing in
  myopic eyes from images obtained by an adaptive optics fundus camera,}
  Japanese Journal of Ophthalmology \textbf{51}, 456--461 (2007).

\bibitem{Li2007}
K.~Y. Li and A.~Roorda, \enquote{Automated identification of cone
  photoreceptors in adaptive optics retinal images,} J. Opt. Soc. Am. A \textbf{24},
  1358--1363 (2007).

\bibitem{Xue2007}
B.~Xue, S.~S. Choi, N.~Doble, and J.~S. Werner, \enquote{Photoreceptor counting
  and montaging of en-face retinal images from an adaptive optics fundus
  camera,} J. Opt. Soc. Am. A \textbf{24}, 1364--1372 (2007).

\bibitem{Chiu2013}
S.~J. Chiu, Y. Lokhnygina, A.~M. Dubis, A. Dubra, J. Carroll, J.~A. Izatt, J.A. and 
  S. Farsiu,\enquote{Automatic cone photoreceptor segmentation using graph 
  theory and dynamic programming,} Biomedical Optics Express \textbf{4 (3)}, 
  924--937 (2013).

\bibitem{Turpin2011}
A.~Turpin, P.~Morrow, B.~Scotney, R.~Anderson, and C.~Wolsley,
  \enquote{Automated identification of photoreceptor cones using multi-scale
  modelling and normalized cross-correlation,} Lecture Notes in Computer
  Science \textbf{6978 LNCS}, 494--503 (2011).

\bibitem{Mohammad2010}
F.~Mohammad, R.~Ansari, J.~Wanek, and M.~Shahidi, \enquote{Photoreceptor cell
  counting in adaptive optics retinal images using content-adaptive filtering,}
  in \emph{Progress in Biomedical Optics and Imaging,} Proc. SPIE \textbf{7626}, 
  76261I (2010).

\bibitem{Loquin2011}
K.~Loquin, I.~Bloch, K.~Nakashima, F.~Rossant, and M.~Paques,
  \enquote{Photoreceptor detection in in-vivo adaptive optics images of the
  retina: Towards a simple interactive tool for the physicians,} in
  \emph{Proceedings - International Symposium on Biomedical Imaging,}
  (IEEE, 2011), pp. 191--194.

\bibitem{Cooper2013}
R.~F. Cooper, C.~S. Langlo, A.~Dubra, and J.~Carroll, \enquote{Automatic
  detection of modal spacing (Yellott's ring) in adaptive optics scanning light
  ophthalmoscope images,} Ophthalmic and Physiological Optics \textbf{33},
  540--549 (2013).
  
\bibitem{Curcio1990}
C.~A. Curcio, K.~R. Sloan, R.~E. Kalina, and A.~E. Hendrickson, \enquote{Human
  photoreceptor topography,} Journal of Comparative Neurology \textbf{292},
  497--523 (1990).

\bibitem{Lombardo2014}
M.~Lombardo, S.~Serrao, and G.~Lombardo, \enquote{Technical Factors 
  Influencing Cone Packing Density Estimates in Adaptive Optics Flood 
  Illuminated Retinal Images,} PloS one \textbf{9(9)}, e107402 (2014).

\bibitem{Chiu2012}
S.~J. Chiu, C.~A. Toth, C.~B. Rickman, J.~A. Izatt, J.A. and S.~Farsiu,
  \enquote{Automatic segmentation of closed-contour features in ophthalmic 
  images using graph theory and dynamic programming,} Biomedical Optics Express 
  \textbf{3 (5)}, 1127--1140 (2012).

\bibitem{Harris1988}
C.~Harris and M.~Stephens, \enquote{A combined corner and edge detector,} in
  \emph{Alvey vision conference,} (1988), vol.~15, p.~50.

\bibitem{Lowe2004}
D.~G. Lowe, \enquote{Distinctive image features from scale-invariant
  keypoints,} International Journal of Computer Vision \textbf{60}, 91--110
  (2004).

\bibitem{Davis1994}
L.~E. Davis, \enquote{A reference guide to the IRAF/DAOPHOT package,} IRAF
  Programming Group, NOAO, Tucson (1994).

\bibitem{Lombardo2012b}
M.~Lombardo, G.~Lombardo, P.~Ducoli, and S.~Serrao, \enquote{Adaptive 
   optics photoreceptor imaging,} Ophthalmology \textbf{119(7)}, 1498--1498 (2012).

\bibitem{Ramaswamy2013}
G.~Ramaswamy and N.~Devaney, \enquote{Pre-processing, registration and
  selection of adaptive optics corrected retinal images,} Ophthalmic and
  Physiological Optics \textbf{33}, 527--539 (2013).
  
\bibitem{Mersereau1979}
R.~M. Mersereau, \enquote{The processing of hexagonally sampled two-dimensional signals,}
  Proceedings of the IEEE \textbf{67 (6)}, 930--949 (1979). 

\bibitem{Park2013}
S.~P. Park, J.~K. Chung, V.~Greenstein, S.~H. Tsang, and S.~Chang, \enquote{A
  study of factors affecting the human cone photoreceptor density measured by
  adaptive optics scanning laser ophthalmoscope,} Experimental Eye Research
  \textbf{108}, 1--9 (2013).

\bibitem{Lombardo2013c}
M.~Lombardo, S.~Serrao, P.~Ducoli, and G.~Lombardo, \enquote{Eccentricity
  dependent changes of density, spacing and packing arrangement of parafoveal
  cones,} Ophthalmic and Physiological Optics \textbf{33}, 516--526 (2013).

\bibitem{Blanco2011}
L.~Blanco, L.~Mugnier, and M.~Glanc, \enquote{Myopic deconvolution of adaptive
  optics retina images,} in \emph{Progress in Biomedical Optics and Imaging,} 
  Proc. SPIE, \textbf{7904}, 790412 (2011).
  
\bibitem{Rubner2000}
Y. Rubner, C. Tomasi and L.~J. Guibas, \enquote{Earth mover's distance as a metric for image retrieval,} 
  International Journal of Computer Vision \textbf{40 (2)}, 99-121 (2000).  

\bibitem{Lee1989}
J.~Lee and K.~Hoppel, \enquote{Noise modeling and estimation of remotely-sensed
  images,} in \emph{Digest - International Geoscience and Remote Sensing 
  Symposium (IGARSS),} (IEEE, 1989), vol.~2, pp. 1005--1008.

\bibitem{Barrett2004}
H.~Barrett and K.~Myers, \emph{Foundations of Image Science}, Wiley series in
  pure and applied optics (Wiley-Interscience, 2004).

\bibitem{Bunch1977}
P.~C. Bunch, J.~F. Hamilton, G.~K. Sanderson, and A.~H. Simmons, \enquote{A
  free response approach to the measurement and characterization of
  radiographic observer performance,} in \emph{Application of Optical
  Instrumentation in Medicine VI} (International Society for Optics and
  Photonics, 1977), pp. 124--135.

\bibitem{Bandos2009}
A.~I. Bandos, H.~E. Rockette, T.~Song, and D.~Gur, \enquote{Area under the
  free-response ROC curve (FROC) and a related summary index,} Biometrics
  \textbf{65}, 247--256 (2009).

\bibitem{Strickland2002}
R.~N. Strickland, \emph{Image-processing techniques for tumor detection} (CRC
  Press, 2002).

\bibitem{Chakraborty1986}
D.~P. Chakraborty, E.~S. Breatnach, M.~V. Yester, B.~Soto, G.~Barnes, and
  R.~Fraser, \enquote{Digital and conventional chest imaging: a modified ROC
  study of observer performance using simulated nodules,} Radiology
  \textbf{158}, 35--39 (1986).

\bibitem{Lombardo2013b}
M.~Lombardo, S.~Serrao, N.~Devaney, M.~Parravano, and G.~Lombardo,
  \enquote{Adaptive optics technology for high-resolution retinal imaging,}
  Sensors (Switzerland) \textbf{13}, 334--366 (2013).
  
\bibitem{Lombardo2014b}
M.~Lombardo, M.~Parravano, G.~Lombardo, M.~Varano, B.~Boccassini, M.~Stirpe, and
  S.~Serrao, \enquote{Adaptive optics imaging of parafoveal cones in type 1 diabetes,}
  Retina \textbf{34 (3)}, 546--557 (2014).
    

\end{thebibliography}
\end{document}